\PassOptionsToPackage{unicode}{hyperref}
\PassOptionsToPackage{hyphens}{url}
\PassOptionsToPackage{dvipsnames,svgnames,x11names}{xcolor}
\documentclass[
]{article}
\usepackage{amsmath,amssymb}
\usepackage{lmodern}
\usepackage{iftex}
\usepackage[T1]{fontenc}
\usepackage[utf8]{inputenc}
\usepackage{textcomp} % provide euro and other symbols
\IfFileExists{upquote.sty}{\usepackage{upquote}}{}
\IfFileExists{microtype.sty}{% use microtype if available
  \usepackage[]{microtype}
  \UseMicrotypeSet[protrusion]{basicmath} % disable protrusion for tt fonts
}{}
\makeatletter
\IfFileExists{parskip.sty}{%
  \usepackage{parskip}
}{% else
  \setlength{\parindent}{0pt}
  \setlength{\parskip}{6pt plus 2pt minus 1pt}}
\makeatother
\usepackage{xcolor}
\NewDocumentCommand\citeproctext{}{}
\NewDocumentCommand\citeproc{mm}{%
  \begingroup\def\citeproctext{#2}\cite{#1}\endgroup}
\makeatletter
 \let\@cite@ofmt\@firstofone
 \def\@biblabel#1{}
 \def\@cite#1#2{{#1\if@tempswa , #2\fi}}
\makeatother
\newlength{\cslhangindent}
\newlength{\csllabelwidth}
\newenvironment{CSLReferences}[2] % #1 hanging-indent, #2 entry-spacing
 {\begin{list}{}{%
  \setlength{\itemindent}{0pt}
  \setlength{\leftmargin}{0pt}
  \setlength{\parsep}{0pt}
  \ifodd #1
   \setlength{\leftmargin}{\cslhangindent}
   \setlength{\itemindent}{-1\cslhangindent}
  \fi
  \setlength{\itemsep}{#2\baselineskip}}}
 {\end{list}}
\usepackage{calc}

\ifLuaTeX
\usepackage[bidi=basic]{babel}
\else
\usepackage[bidi=default]{babel}
\fi
\babelprovide[main,import]{american}
\def\languageshorthands#1{}
\IfFileExists{bookmark.sty}{\usepackage{bookmark}}{\usepackage{hyperref}}
\IfFileExists{xurl.sty}{\usepackage{xurl}}{} % add URL line breaks if available
\hypersetup{
  pdftitle={RydIQule Version 2: Enhancing graph-based modeling of
Rydberg atoms},
  pdfauthor={Benjamin N. Miller, David H. Meyer, Carter A. Montag, Omar
Nagib, Teemu Virtanen, Peter K. Elgee, Kevin C. Cox},
  pdflang={en-US},
  colorlinks=true,
  linkcolor={Maroon},
  filecolor={Maroon},
  citecolor={Blue},
  urlcolor={Blue},
  pdfcreator={LaTeX via pandoc}}

\title{RydIQule Version 2: Enhancing graph-based modeling of Rydberg
atoms}

\definecolor{c53baa1}{RGB}{83,186,161}
\definecolor{c202826}{RGB}{32,40,38}
\def \rorglobalscale {0.1}
\newcommand{\rorlogo}{%
\begin{tikzpicture}[y=1cm, x=1cm, yscale=\rorglobalscale,xscale=\rorglobalscale, every node/.append style={scale=\rorglobalscale}, inner sep=0pt, outer sep=0pt]
  \begin{scope}[even odd rule,line join=round,miter limit=2.0,shift={(-0.025, 0.0216)}]
    \path[fill=c53baa1,nonzero rule,line join=round,miter limit=2.0] (1.8164, 3.012) -- (1.4954, 2.5204) -- (1.1742, 3.012) -- (1.8164, 3.012) -- cycle;
    \path[fill=c53baa1,nonzero rule,line join=round,miter limit=2.0] (3.1594, 3.012) -- (2.8385, 2.5204) -- (2.5172, 3.012) -- (3.1594, 3.012) -- cycle;
    \path[fill=c53baa1,nonzero rule,line join=round,miter limit=2.0] (1.1742, 0.0669) -- (1.4954, 0.5588) -- (1.8164, 0.0669) -- (1.1742, 0.0669) -- cycle;
    \path[fill=c53baa1,nonzero rule,line join=round,miter limit=2.0] (2.5172, 0.0669) -- (2.8385, 0.5588) -- (3.1594, 0.0669) -- (2.5172, 0.0669) -- cycle;
    \path[fill=c202826,nonzero rule,line join=round,miter limit=2.0] (3.8505, 1.4364).. controls (3.9643, 1.4576) and (4.0508, 1.5081) .. (4.1098, 1.5878).. controls (4.169, 1.6674) and (4.1984, 1.7642) .. (4.1984, 1.8777).. controls (4.1984, 1.9719) and (4.182, 2.0503) .. (4.1495, 2.1132).. controls (4.1169, 2.1762) and (4.0727, 2.2262) .. (4.0174, 2.2635).. controls (3.9621, 2.3006) and (3.8976, 2.3273) .. (3.824, 2.3432).. controls (3.7505, 2.359) and (3.6727, 2.367) .. (3.5909, 2.367) -- (2.9676, 2.367) -- (2.9676, 1.8688).. controls (2.9625, 1.8833) and (2.9572, 1.8976) .. (2.9514, 1.9119).. controls (2.9083, 2.0164) and (2.848, 2.1056) .. (2.7705, 2.1791).. controls (2.6929, 2.2527) and (2.6014, 2.3093) .. (2.495, 2.3487).. controls (2.3889, 2.3881) and (2.2728, 2.408) .. (2.1468, 2.408).. controls (2.0209, 2.408) and (1.905, 2.3881) .. (1.7986, 2.3487).. controls (1.6925, 2.3093) and (1.6007, 2.2527) .. (1.5232, 2.1791).. controls (1.4539, 2.1132) and (1.3983, 2.0346) .. (1.3565, 1.9436).. controls (1.3504, 2.009) and (1.3351, 2.0656) .. (1.3105, 2.1132).. controls (1.2779, 2.1762) and (1.2338, 2.2262) .. (1.1785, 2.2635).. controls (1.1232, 2.3006) and (1.0586, 2.3273) .. (0.985, 2.3432).. controls (0.9115, 2.359) and (0.8337, 2.367) .. (0.7519, 2.367) -- (0.1289, 2.367) -- (0.1289, 0.7562) -- (0.4837, 0.7562) -- (0.4837, 1.4002) -- (0.6588, 1.4002) -- (0.9956, 0.7562) -- (1.4211, 0.7562) -- (1.0118, 1.4364).. controls (1.1255, 1.4576) and (1.2121, 1.5081) .. (1.2711, 1.5878).. controls (1.2737, 1.5915) and (1.2761, 1.5954) .. (1.2787, 1.5991).. controls (1.2782, 1.5867) and (1.2779, 1.5743) .. (1.2779, 1.5616).. controls (1.2779, 1.4327) and (1.2996, 1.3158) .. (1.3428, 1.2113).. controls (1.3859, 1.1068) and (1.4462, 1.0176) .. (1.5237, 0.944).. controls (1.601, 0.8705) and (1.6928, 0.8139) .. (1.7992, 0.7744).. controls (1.9053, 0.735) and (2.0214, 0.7152) .. (2.1474, 0.7152).. controls (2.2733, 0.7152) and (2.3892, 0.735) .. (2.4956, 0.7744).. controls (2.6016, 0.8139) and (2.6935, 0.8705) .. (2.771, 0.944).. controls (2.8482, 1.0176) and (2.9086, 1.1068) .. (2.952, 1.2113).. controls (2.9578, 1.2253) and (2.9631, 1.2398) .. (2.9681, 1.2544) -- (2.9681, 0.7562) -- (3.3229, 0.7562) -- (3.3229, 1.4002) -- (3.4981, 1.4002) -- (3.8349, 0.7562) -- (4.2603, 0.7562) -- (3.8505, 1.4364) -- cycle(0.9628, 1.7777).. controls (0.9438, 1.7534) and (0.92, 1.7357) .. (0.8911, 1.7243).. controls (0.8623, 1.7129) and (0.83, 1.706) .. (0.7945, 1.7039).. controls (0.7588, 1.7015) and (0.7252, 1.7005) .. (0.6932, 1.7005) -- (0.4839, 1.7005) -- (0.4839, 2.0667) -- (0.716, 2.0667).. controls (0.7477, 2.0667) and (0.7805, 2.0643) .. (0.8139, 2.0598).. controls (0.8472, 2.0553) and (0.8768, 2.0466) .. (0.9025, 2.0336).. controls (0.9282, 2.0206) and (0.9496, 2.0021) .. (0.9663, 1.9778).. controls (0.9829, 1.9534) and (0.9914, 1.9209) .. (0.9914, 1.8799).. controls (0.9914, 1.8362) and (0.9819, 1.8021) .. (0.9628, 1.7777) -- cycle(2.6125, 1.3533).. controls (2.5889, 1.2904) and (2.5553, 1.2359) .. (2.5112, 1.1896).. controls (2.4672, 1.1433) and (2.4146, 1.1073) .. (2.3529, 1.0814).. controls (2.2916, 1.0554) and (2.2228, 1.0427) .. (2.1471, 1.0427).. controls (2.0712, 1.0427) and (2.0026, 1.0557) .. (1.9412, 1.0814).. controls (1.8799, 1.107) and (1.8272, 1.1433) .. (1.783, 1.1896).. controls (1.7391, 1.2359) and (1.7052, 1.2904) .. (1.6817, 1.3533).. controls (1.6581, 1.4163) and (1.6465, 1.4856) .. (1.6465, 1.5616).. controls (1.6465, 1.6359) and (1.6581, 1.705) .. (1.6817, 1.7687).. controls (1.7052, 1.8325) and (1.7388, 1.8873) .. (1.783, 1.9336).. controls (1.8269, 1.9799) and (1.8796, 2.0159) .. (1.9412, 2.0418).. controls (2.0026, 2.0675) and (2.0712, 2.0804) .. (2.1471, 2.0804).. controls (2.223, 2.0804) and (2.2916, 2.0675) .. (2.3529, 2.0418).. controls (2.4143, 2.0161) and (2.467, 1.9799) .. (2.5112, 1.9336).. controls (2.5551, 1.8873) and (2.5889, 1.8322) .. (2.6125, 1.7687).. controls (2.636, 1.705) and (2.6477, 1.6359) .. (2.6477, 1.5616).. controls (2.6477, 1.4856) and (2.636, 1.4163) .. (2.6125, 1.3533) -- cycle(3.8015, 1.7777).. controls (3.7825, 1.7534) and (3.7587, 1.7357) .. (3.7298, 1.7243).. controls (3.701, 1.7129) and (3.6687, 1.706) .. (3.6333, 1.7039).. controls (3.5975, 1.7015) and (3.5639, 1.7005) .. (3.5319, 1.7005) -- (3.3226, 1.7005) -- (3.3226, 2.0667) -- (3.5547, 2.0667).. controls (3.5864, 2.0667) and (3.6192, 2.0643) .. (3.6526, 2.0598).. controls (3.6859, 2.0553) and (3.7155, 2.0466) .. (3.7412, 2.0336).. controls (3.7669, 2.0206) and (3.7883, 2.0021) .. (3.805, 1.9778).. controls (3.8216, 1.9534) and (3.8301, 1.9209) .. (3.8301, 1.8799).. controls (3.8301, 1.8362) and (3.8206, 1.8021) .. (3.8015, 1.7777) -- cycle;
  \end{scope}
\end{tikzpicture}
}

\usepackage[affil-it]{authblk}
\usepackage{orcidlink}
\author[1%
  *%
  ]{Benjamin N. Miller%
    \,\orcidlink{0000-0003-0017-1355}\,%
    }
\author[1%
  *%
  ]{David H. Meyer%
    \,\orcidlink{0000-0003-2452-2017}\,%
    }
\author[2%
  ]{Carter A. Montag%
    \,\orcidlink{0009-0000-9583-3854}\,%
    }
\author[3%
  ]{Omar Nagib%
    \,\orcidlink{0000-0003-3345-4781}\,%
    }
\author[4%
  ]{Teemu Virtanen%
    \,\orcidlink{0000-0002-0653-2564}\,%
    }
\author[1%
  ]{Peter K. Elgee%
    \,\orcidlink{0000-0002-5055-3039}\,%
    }
\author[1%
  ]{Kevin C. Cox%
    \,\orcidlink{0000-0001-5049-3999}\,%
    }

\affil[1]{DEVCOM Army Research Laboratory, 2800 Powder Mill Rd, Adelphi,
MD, 20783, USA%
    \,\protect\href{https://ror.org/011hc8f90}{\protect\rorlogo}\,%
  }
\affil[2]{Program in Applied Mathematics, University of Arizona, 1200 E
University Blvd, Tuscon, AZ 85721, USA%
    \,\protect\href{https://ror.org/03m2x1q45}{\protect\rorlogo}\,%
  }
\affil[3]{Department of Physics, University of Wisconsin-Madison, 1150
University Avenue, Madison, WI, 53706, USA%
    \,\protect\href{https://ror.org/01y2jtd41}{\protect\rorlogo}\,%
  }
\affil[4]{Naval Air Warfare Center, 1 Administration Circle, China Lake,
CA, 93555, USA%
    \,\protect\href{https://ror.org/01f0pxq13}{\protect\rorlogo}\,%
  }
\affil[*]{These authors contributed equally.}
\date{13 March 2025}

\begin{document}
\maketitle

\section{Summary}\label{summary}

Rydberg atomic radio-frequency (rf) sensors are an emerging technology
platform that relies on vaporous atoms, interrogated with laser beams
and nearly ionized, to receive rf signals. Rydberg rf sensors have a
number of interesting fundamental distinctions from traditional receiver
technologies, such as those based on metallic antennas, since they are
governed by the quantum physics of atom-light interactions
(\citeproc{ref-fancher_rydberg_2021}{Fancher et al., 2021}). As Rydberg
sensors quickly advance from laboratory experiments into fieldable
devices, there is a need for a general software modeling tool that fully
encompasses the internal physics of the sensor. The Rydberg Interactive
Quantum Module (RydIQule) is a Python package designed to fill this
need.

RydIQule calculates the dynamics of atomic quantum states, with a
particular emphasis on modeling thermal vapors of Rydberg atoms coupled
to optical and radio-frequency electromagnetic fields. To accomplish
this, a unique graph-based paradigm is used to represent the complex
quantum system consisting of multiple energy levels and multiple
electromagnetic fields, where each energy level is stored as a graph
node and each electromagnetic coupling as a graph edge. RydIQule then
generates a set of differential equations for the quantum state dynamics
from the graph, using the Lindblad master equation formalism
(\citeproc{ref-manzano_short_2020}{Manzano, 2020}). Finally, RydIQule
leverages linear equation solvers, such as those provided by NumPy
(\citeproc{ref-harris_array_2020}{Harris et al., 2020}), SciPy
(\citeproc{ref-virtanen_scipy_2020}{Virtanen et al., 2020}) or CyRK
(\citeproc{ref-cyrk}{Renaud, 2022}) to efficiently solve these systems
and recover the quantum system response to arbitrary input fields.
During the numerical solving, systems of equations are represented as
tensor objects, allowing for efficient parameterization and computation
of large sets of equations. All together, RydIQule provides a flexible
platform for forward modeling Rydberg sensors while also providing a
widely useful set of theoretical tools for fundamental exploration of
atomic physics concepts.

The initial public release of RydIQule in late 2023 built the core
functionality described above
(\citeproc{ref-miller_rydiqule_2024}{Miller et al., 2024}). Here we
outline RydIQule's version 2 release which expands on its capabilities
to more accurately model real-world atoms.

\section{Statement of Need}\label{statement-of-need}

The unique quantum properties of Rydberg atoms offer distinct advantages
in the fields of sensing, communication, and quantum information
(\citeproc{ref-adams_rydberg_2019}{Adams et al., 2019}). However, the
breadth of possible configurations and experimental parameters makes
general modeling of an experiment difficult. One challenge is that many
atomic energy levels consist of numerous magnetic sublevels that arise
from the different possible orientations of the electron's and nucleus's
angular momentum. These sublevels have different responses to applied
magnetic and electric fields which leads to measureable differences for
most real-world atomic sensors. In some cases, this sublevel structure
can be treated in average and safely ignored. More often however, they
are ignored due to the significant complexity inherent in expanding the
model size to account for them.

For example, in accounting for magnetic splitting, a typical Rydberg
spectroscopy experiment using the
\(5\text{S}_{1/2}\rightarrow5\text{P}_{3/2}\rightarrow n\text{D}\) set
of transitions would have a total of 46 levels with up to 34
dipole-allowed couplings between them. In RydIQule's initial release,
users would have no choice but to individually add each sublevel and the
many associated electromagnetic couplings, making it of little
functional use. For this reason, RydIQule was not easily scaled to
realistic scenarios involving several atomic states and typically many
tens, or possibly even hundreds, of sublevels.

The main advance of RydIQule version 2 is to allow user-friendly
inclusion of large atomic manifolds that include the complete set of
electronic and magnetic sublevels. In particular, this release
introduces a new paradigm for structured labeling of states using
arbitrary tuples, and expands the automated calculation of relevant
atomic properties on alkali atoms commonly used in Rydberg physics to
include sublevels. This release also includes a new steady-state
Doppler-averaging method that greatly improves speed and accuracy, along
with many other optimizations and improvements to the code-base.

\subsection{Handling Sublevel
Structure}\label{handling-sublevel-structure}

RydIQule's primary improvement in version 2 is in handling state
manifolds: degenerate sets of sublevel states defined by a magnetic
interaction with the electron. It handles this sublevel structure by
expanding the way nodes are labelled. Rather than only using integers,
arbitrary tuples can now be used as graph nodes. This allows for
manifolds to be defined by using tuples in a way that directly maps to
the atomic structure. RydIQule's core functions relating to graph
operations have been updated to interchangeably address individual
states or entire manifolds. It's internals have been overhauled to not
only ensure that all relevant states/couplings are added, but tracked as
originating from a single manifold.

\subsection{Improved Calculation of Atomic
Properties}\label{improved-calculation-of-atomic-properties}

Version 2 of RydIQule also completely overhauls the \texttt{Cell} class,
which provides automatic calculation of atomic properties of alkali
atoms using the Alkali.ne Rydberg Calculator (ARC) package
(\citeproc{ref-robertson_arc_2021}{Robertson et al., 2021};
\citeproc{ref-sibalic_arc_2017}{Šibalić et al., 2017}). In version 1,
this class could only handle simplified atomic models that treated
manifolds of atomic sublevels as a single approximate state. Though this
type of model is very fast and can be effective in many situations, it
breaks down for systems in the presence of magnetic fields (including
those as weak as Earth's background magnetic field) or for large
electric field amplitudes that result in inhomogeneous couplings due to
sublevel structure. By leveraging the tuple labelling outlined above,
\texttt{Cell} can now define states by their quantum numbers directly,
which allows for natural definition and coupling of entire manifolds.

Version 2 also greatly enhances the leveraging of ARC to calculate more
system parameters automatically. In particular, there is automatic
calculation of coupling strengths between manifolds defined in
incommensurate fine and hyperfine bases. This feature allows for more
efficient modeling of Rydberg atoms since low energy and high energy
states can be defined in their natural bases, fine and hyperfine
respectively, lowering the total number of sublevels that need to be
calculated.

\subsection{Analytic Doppler
Averaging}\label{analytic-doppler-averaging}

Support for steady-state Doppler-averaged models leveraging an exact
analytic method has been added. This functionality is based on the
theoretical work presented in (\citeproc{ref-nagib_exact_2025}{Nagib \&
Walker, 2025}). That work derives the exact velocity dependence due to
Doppler shifts for a system along a single axis. It effectively reduces
the Doppler-averaging integration along that single dimension to two
diagonalizations, avoiding velocity sampling that dimension, and
enabling a general, analytic result. For example, a two-dimensional
Doppler-average only needs to be numerically sampled along one axis,
with the other performed analytically. This reduction in dimensionality
results in over an order of magnitude reduction in calculation time and
memory footprint while returning significantly higher accuracy
solutions. Experimental support for 1D solves only was released in
version 2.0.0, with RydIQule v2.1.0 providing full support.

\section{Related Packages and Work}\label{related-packages-and-work}

Modeling quantum systems using the semi-classical Lindblad formalism is
a common task that has been implemented by many physicists for their
bespoke problems. Other tools that implement this type of simulation for
specific types of problems include: qubits in QuTiP
(\citeproc{ref-johansson_qutip_2013}{Johansson et al., 2013}), atomic
magnetometers in ElecSus (\citeproc{ref-keaveney_elecsus_2018}{Keaveney
et al., 2018}), and laser cooling in PyLCP
(\citeproc{ref-eckel_pylcp_2022}{Eckel et al., 2022}). Ultimately, the
goal of RydIQule has not been to develop a new modeling technique, but
rather to make a common, flexible, and most importantly efficient tool
that solves a ubiquitous problem.

RydIQule's version 2 release aims to capture the functionality of the
Atomic Density Matrix (ADM) package
(\citeproc{ref-rochester_atomicdensitymatrix_2008}{Rochester, 2008})
written in Mathematica. While very capable, it is built on a proprietary
platform requiring a paid license which limits its accessibility. And
since Mathematica is an interpreted language, it can lack the speed that
compiled libraries like NumPy enable, especially when exploring a large
parameter space.

Since RydIQule version 1 has been publicly released, it has been used in
several publications to model both general Rydberg atom physics
(\citeproc{ref-backes_performance_2024}{Backes et al., 2024};
\citeproc{ref-glick_warm_2025}{Glick et al., 2025};
\citeproc{ref-su_two-photon_2025}{Su et al., 2025}) as well as Rydberg
sensor development (\citeproc{ref-cui_realizing_2025}{Cui et al., 2025};
\citeproc{ref-elgee_satellite_2023}{Elgee et al., 2023};
\citeproc{ref-gokhale_deep_2024}{Gokhale et al., 2024};
\citeproc{ref-richardson_study_2025}{Richardson et al., 2025};
\citeproc{ref-santamaria-botello_comparison_2022}{Santamaria-Botello et
al., 2022}).

\section{Acknowledgements}\label{acknowledgements}

Financial support for the development of RydIQule version 2 was provided
by the DEVCOM Army Research Laboratory. Carter Montag acknowledges
financial support from the Department of Defense Historically Black
Colleges \& Universities and Minority Serving Institutions (HBCU/MI)
Summer Research Internship Program. Omar Nagib acknowledges helpful
discussion with Thad G. Walker. The views, opinions and/or findings
expressed are those of the authors and should not be interpreted as
representing the official views or policies of the Department of Defense
or the U.S. Government.

\phantomsection\label{refs}

\end{document}